Wafer-scale fabrication of 2D nanostructures via thermomechanical nanomolding

Mehrdad T Kiani[a], Quynh P Sam[a], Yeon Sik Jung[b], Hyeuk Jin Han[c], Judy J Cha[a*]

a) Department of Materials Science and Engineering, Cornell University, Ithaca, NY
b) Department of Materials Science and Engineering, Korea Advanced Institute of Science and Technology, Daejeon, South Korea
c) Department of Environment and Energy Engineering, Sungshin Women's University, Seoul, South Korea

* corresponding author: judy.cha@cornell.edu

**Abstract**

With shrinking dimensions in integrated circuits, sensors, and functional devices, there is a pressing need to develop nanofabrication techniques with simultaneous control of morphology, microstructure, and material composition over wafer length scales. Current techniques are largely unable to meet all these conditions, suffering from poor control of morphology and defect structure or requiring extensive optimization or post-processing to achieve desired nanostructures. Recently, thermomechanical nanomolding (TMNM) has been shown to yield single-crystalline, high aspect ratio nanowires of metals, alloys, and intermetallics over wafer-scale distances. Here, we extend TMNM for wafer-scale fabrication of 2D nanostructures. Using Cu, we successfully nanomold Cu nanoribbons with widths < 50 nm, depths ~ 0.5-1 µm and lengths ~ 7 mm into Si trenches at conditions compatible with back end of line processing. Through SEM cross-section imaging and 4D-STEM grain orientation maps, we show that the grain size of the bulk feedstock is transferred to the nanomolded structures up to and including single crystal Cu. Based on the retained microstructures of molded 2D Cu, we discuss the deformation mechanism during molding for 2D TMNM.

**Introduction**

The ability to fabricate increasingly complex nanostructures is the backbone of numerous advances in wide ranging fields including catalysis[1], energy storage[2], integrated circuits[3], plasmonics[4], and biomedical diagnostics[5]. However, a nanofabrication approach that enables simultaneous control over crystallinity, size, morphology, and material composition remains elusive. Bottom-up techniques such as molecular beam epitaxy and atomic layer deposition can fabricate exceptionally high quality single crystalline films and coatings but largely fail at fabrication of other more complex nanostructures[6,7]. Further, these techniques do not readily translate to different material classes, require extensive optimization for each material composition of interest, and control of grain size and orientation is limited or dictated by the underlying substrate. Colloidal synthesis can achieve excellent size, shape, and composition control but this technique is limited to 0D or 1D nanostructures, requires surface ligands to stabilize in solution, and cannot be easily integrated with other nanofabrication techniques or cleanroom processing[8].

Micro-additive manufacturing, such as two-photon lithography or electrohydrodynamic redox 3D printing, enables arbitrary control of morphology and rapid optimization but is limited to few material classes, cannot be performed over large areas due to the limitations of the rastering of a laser or a tip, and the final structures are nanoporous[9,10]. Top-down techniques, such as lithography, enable control of morphology and size but are extremely limited in material choice and lack control of crystallinity.

Recently, thermomechanical nanomolding (TMNM), a technique whereby a bulk feedstock of a desired material is pressed through a nanoporous mold at elevated temperatures and pressures, has been shown to yield high aspect ratio single crystal nanowires over wafer-scale distances[11,12]. TMNM is material-agnostic, having been shown to work for a wide variety of materials including metals[11,13], alloys[11], and intermetallics[14,15]. TMNM of 1D nanowires relies on initial grain reorientation near the base of the nanowire as the bulk feedstock enters the nanopores of the mold to the preferred growth direction, which minimizes total surface energy and interfacial diffusion barrier along the mold walls[12]. While 1D TMNM can form nanowires of a desired material, most applications require the ability to fabricate more complex and higher dimensional nanostructures. At present, it is unknown whether interfacial diffusion will still occur during TMNM to form nanostructures beyond nanowires, which have different boundary conditions and surface area to volume ratios, and if additional growth mechanisms are at play.

Here, we use TMNM to form 2D Cu nanostructures inside $Si_3N_4$/Si trenches. Using Cu feedstocks with varying grain sizes, we observe successful TMNM over millimeter length scales of 2D nanostructures with high aspect ratios up to 50. Importantly, the molded 2D Cu nanostructures retain the same grain size of the Cu feedstock, from nanoscale grains to single crystal. The conditions used for TMNM are back end of line (BEOL) compatible and scalable, enabling future integration with semiconductor processing.

**Results**

TMNM in 2D requires a bulk feedstock of the material of interest along with a mold (Figure 1a) pressed together at elevated temperature and pressure. We use 2D Si trenches fabricated using traditional lithographic techniques as the mold for TMNM (Figure 1b). The trenches have either a width of ~20 or ~40 nm with depths of 500 or 1000 nm. These trenches have a ~100 nm thick $Si_3N_4$ barrier layer to prevent silicide formation at elevated temperatures. For each molding experiment, a 7 mm x 7 mm size wafer is used. The experimental setup is shown in Figure 1a. For the experiments, three Cu samples were used with different grain sizes: nanocrystalline Cu (nc-Cu) with grain sizes ranging between 50 nm - 7 µm (Figure 1c), microcrystalline Cu (µc-Cu) with grain sizes ranging from 1 - 60 µm (Figure 1d), and a <110>-oriented single crystal Cu (sc-Cu). All TMNM was done in an argon environment to prevent Cu oxidation during molding at elevated temperatures. A pressure of 30 MPa was used for TMNM of nc-Cu foils and 60 MPa for µc-Cu. These pressures did not lead to crack formation in the Si wafer. After TMNM, the remaining Cu foil was removed manually using tweezers and no further processing or cleaning was performed.

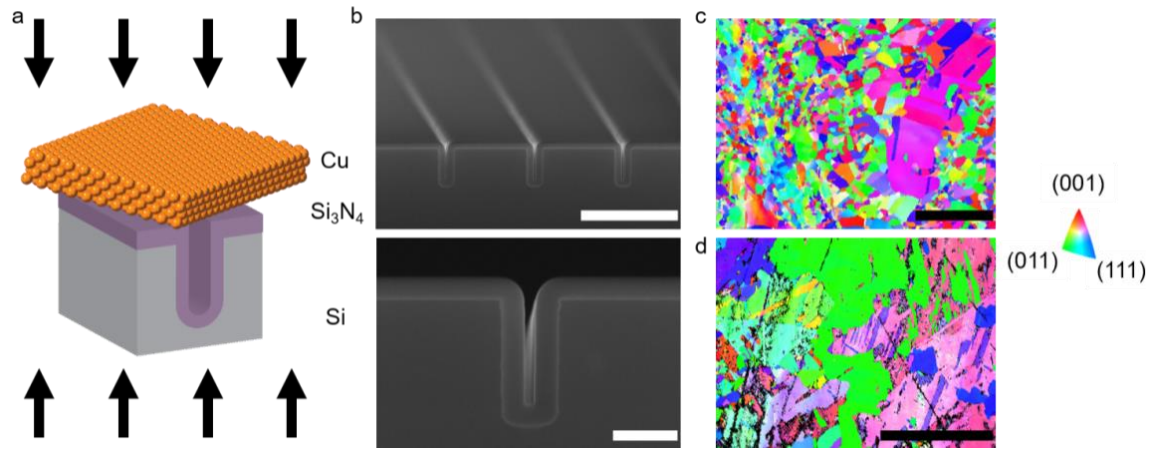

Figure 1. Setup of TMNM for Cu molding. a) Schematic of 2D TMNM setup. Orange denotes Cu crystal, purple denotes $Si_3N_4$ barrier layer, and grey indicates Si wafer. b) SEM images of the $Si_3N_4$/Si mold. Scale bar is 1 µm and 100 nm. c) Electron backscatter diffraction (EBSD) map of nanocrystalline copper. Scale bar is 5 µm. d) EBSD map of microcrystalline copper. Scale bar is 100 µm.

Successful 2D TMNM of the nc-Cu foil is observed using an optical microscope over an extremely large area (Figure 2a). An optical image of a larger region is shown in Supplementary Figure 1. Cross-section SEM imaging shows complete molding with Cu thoroughly filling the trenches to form Cu nanoribbons (Figure 2b). SEM imaging along the trench shows the grain size and morphology (Figure 2c). The extent of TMNM is both time and temperature dependent. At 200°C ($0.35T_m$ where $T_m$ is the melting temperature of bulk Cu), TMNM was not successful. At 300°C ($0.42T_m$) and 60 minutes, TMNM produces incomplete molding where gaps remain between the bottom of the trench and molded Cu. These gaps are found at the interface between Cu grains with sharp angles (Figure 2c, top). Increasing the temperature to 400°C ($0.5T_m$) led to more complete nanomolding with minimal gaps near the end of the mold (Figure 2c, middle). 400°C and 90 minutes was determined to be the ideal nanomolding conditions for nc-Cu (Figure 2c, bottom and Supplementary Figure 2). Whereas 1D TMNM always led to single crystal nanowires[13], we observe that 2D TMNM of nc-Cu led to the nanocrystalline grains and twins in the molded Cu ribbons with comparable sizes to the nc-Cu feedstock. The grain shape is random, with no directional or columnar grain structure, which suggests no preferred grain reorientation occurs during 2D TMNM. We also successfully nanomolded large aspect ratio Cu ribbons with a width of 20 nm and depth of 1 µm (Figure 2e) at 400°C, 30 MPa, and 90 minutes with nc-Cu. There is no difference in grain shape or structure with the increased aspect ratio.

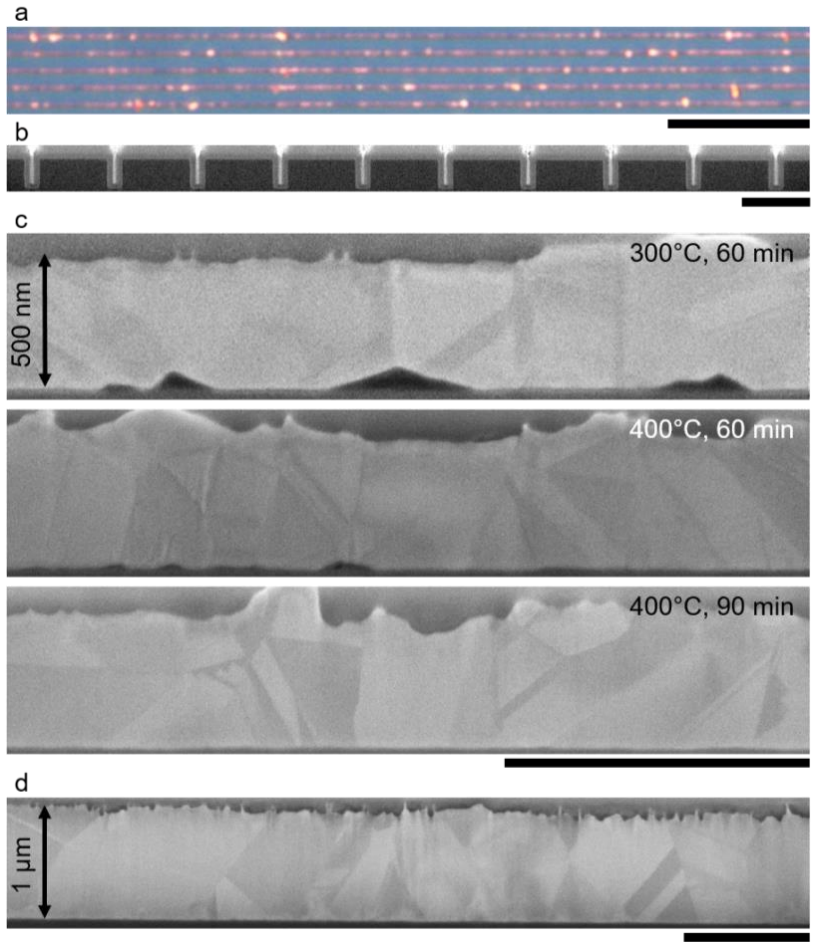

Figure 2. TMNM of nc-Cu. a) Optical bright field image of nc-Cu filled trenches. Scale bar is 10 µm. b) SEM cross-section image of nc-Cu filled trenches. Scale bar is 1 µm. c) SEM cross section images along nc-Cu filled trenches for increasing temperature and time. Scale bar is 1 µm. d) High aspect ratio trenches filled with nc-Cu. Scale bar is 1 µm.

Scanning transmission electron microscopy (STEM) imaging of liftout samples of Cu-filled trenches shows no apparent intermixing of Cu and Si at the Cu/ $Si_3N_4$ interface and complete molding for nc-Cu at 400°C and 90 minutes (Figure 3a). 4D-STEM maps of the nanomolded trenches were collected to determine grain orientation of the molded nc-Cu in the trenches using electron diffraction patterns[16] (Figure 3a). For the nc-Cu foil, the molded Cu show a variety of grain structures, ranging from high angle grain boundaries (Figure 3b) to low angle grain boundaries and single crystal trenches (Figure 3c). In the low angle grain boundary case, plotting grain boundary misorientation angle along the nanomolding direction shows a gradual rotation in the crystal before the grain boundary, whereas intragranular rotations in the single crystal case is random (Figure 3c). The polycrystalline nature is also observed in a partially molded nc-Cu trench at 300°C, 30 MPa, and 90 minutes, indicating that the grains are present during the entire nanomolding process (Figure 3d). A variety of grain structures were also seen for nanomolded nc-Cu at 300°C, 30 MPa, and 90 minutes (Supplementary Figure 3).

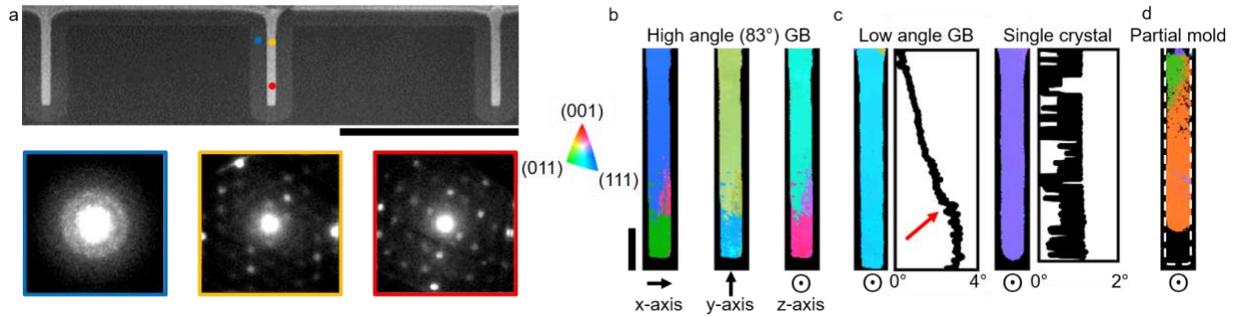

Figure 3. Grain analysis of molded nc-Cu. a) STEM imaging of nc-Cu filled trenches with representative electron diffraction patterns taken from three locations, indicated with colored dots, and the Si trench mold with $Si_3N_4$ lining. Scale bar is 100 nm. B) 4D-STEM grain orientation maps of molded nc-Cu shown in a). Scale bar is 100 nm. C) 4D-STEM grain orientation map of additional Cu-filled trenches. Red arrow indicates approximate position of low angle grain boundary. d) Grain orientation map of nc-Cu sample partially molded at 300°C, 30 MPa, and 90 minutes. White dotted line denotes the $Si_3N_4$/Si trench boundary.

A higher pressure of 60 MPa was needed to achieve successful nanomolding for µc-Cu. Surprisingly, when TMNM was performed with µc-Cu foil at 400°C and for 90 minutes, we see little intensity contrast in SEM cross-section images and larger grain sizes compared to nc-Cu (Figure 4a). Thus, the molded Cu retains the large grains of the original µc-Cu feedstock. At the single crystal limit using <110>-oriented Cu, nanomolding was strongly dependent on the orientation of the crystal with respect to the trench walls. When {111} planes were not parallel to the trench wall, nanomolding at 400°C and ~100 MPa was unsuccessful. Higher pressures were not feasible since they led to extensive cracking in the Si wafer. When Cu {111} planes were aligned parallel to the trench wall, nanomolding at 400°C, 70 MPa, and 90 minutes was successful (Figure 4b). In addition to Cu trench formation, we also form freestanding 2D Cu fins (Figure 4c). Here, we use the sc-Cu as the feedstock and etch away the Si and $Si_3N_4$ after molding. 4D-STEM grain orientation maps of the Cu fins show that the orientation of the fins matches the orientation of the bulk feedstock with the <111> direction aligned with the trench wall, the <211> direction aligned along the length of the trench, and the <110> direction aligned with the nanomolding direction (Figure 4d, Figure 4e, Supplementary Figure 4, and Supplementary Figure 5).

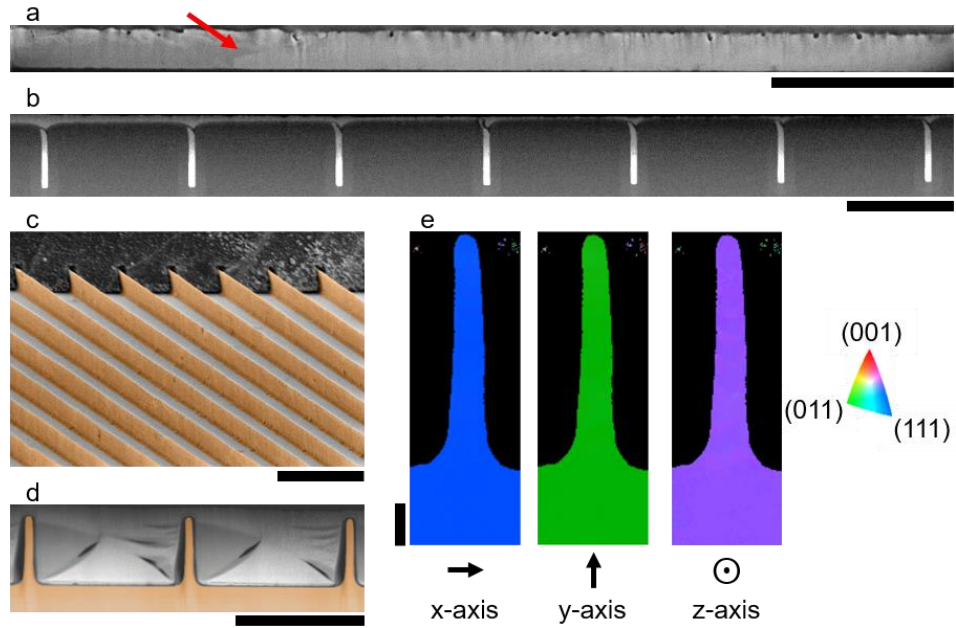

Figure 4. TMNM of µc-Cu and sc-Cu. a) SEM cross-section imaging of µc-Cu filled trench. Red arrow denotes grain boundary. Scale bar is 2 µm. b) STEM image of sc-Cu filled trench. Scale bar is 1 µm. c) False color SEM image of sc-Cu nanofins. Scale bar is 2 µm. d) False color STEM image of Cu nanofins. Scale bar is 1 µm. e) 4D STEM grain orientation map of representative Cu nanofin. Scale bar is 100 nm.

**Discussion on diffusion mechanism in 2D TMNM**

The different microstructures observed in nanomolded Cu enable us to identify the potential growth mechanisms for TMNM in 2D. As with 1D, 2D TMNM is both time and temperature dependent with increased temperature promoting more complete molding. There is a slight pressure increase necessary to mold nc-Cu (30 MPa), µc-Cu (60 MPa), and sc-Cu (70 MPa). The lack of any consistent grain orientation in nc-Cu, the presence of grains in partially molded samples, and comparable grain size in molded Cu with grain size in the bulk feedstock indicate that the microstructure of the feedstock is conserved during molding. Finally, sc-Cu was only successfully nanomolded when {111} planes were aligned with the walls of the trench, matching what is expected from 1D TMNM.

Here, we discuss differences between 1D and 2D TMNM. TMNM in 1D is driven by crystal reorientation at the base of the nanowire to minimize surface energy followed by interfacial diffusion along the mold wall[11]. In the case of FCC metals such as Cu, the preferred molding direction is <110>, which maximizes the area of {111} planes on the nanowire surface since it has the lowest surface energy. If crystal reorientation occurred in the trenches to minimize surface energy for TMNM in 2D, we expect preferred grain orientations that would lower the surface energy of the molded nc-Cu samples, which we did not observe. We estimate the surface energy for the planes facing the trench wall from the orientation maps generated from 4D-STEM for the nc-Cu samples both at 300°C and 400°C. For both cases, there is no clear relation between the surface energy of the entrance grain and the final grain at the bottom of trench. For example, for the trench shown in Figure 3b, the grain near the mold entrance has ~(543) planes aligned with the trench wall, which have a surface energy of ~1.06 J/m² [17]. The grain at the bottom of the trench has ~(771) aligned with the trench wall, which have a slightly higher surface energy of ~1.1 J/m² [17]. Similarly, the single crystal grain shown in Figure 3c has ~(421) planes aligned with the trench walls, which has a high surface energy of 1.14 J/m², yet we observe no crystal rotation[17]. Thus, while {111} planes were highly preferred in the single crystal Cu molding (Figure 4), polycrystalline bulk feedstocks showed no clear preference as to grain orientation during molding.

The lack of grain rotation in the case of 2D TMNM in contrast to 1D TMNM can partially be attributed to different boundary conditions during molding. In 1D TMNM, there is a high probability for a random grain to have a set of {111} planes oriented on the mold walls as the mold wall surrounds the grain. In 1D TMNM, once the grain has the <110>-direction aligned with the molding direction, there will always be {111} planes aligned with the mold walls. In 2D nanomolding, the grain must be oriented such that the <110>-direction is in the molding direction and {111} planes align with the trench wall. Thus, 2D TMNM requires more grain rotation for proper alignment, which could become a large energetic barrier. Experimentally, this was observed for the single crystal Cu molding where misaligned crystals did not mold even at high pressures of ~100 MPa.

We instead attribute the resulting microstructure observed in 2D TMNM to grain boundary diffusion and subsequent grain boundary sliding resulting from diffusion. The predicted length of the molded nanowire in 1D TMNM, $L$, is shown in Equation 1[11]:

$$L = \sqrt{L_0^2 + 2\frac{p\Omega t}{k_B T}\frac{DS}{A}} \quad (1)$$

where $L_0$ is related to the entry effect during molding, $p$ is the pressure, $D$ is the diffusion constant, $\Omega$ is the atomic volume, $t$ is the molding time, $k_b$ is Boltzmann constant, $T$ is the temperature, $S$ is the cross-section area of the diffusion path, and $A$ is the cross-section area of the mold. In 1D TMNM, interfacial diffusion of Cu on the mold surface drives formation of nanowires. For 2D TMNM, we include the contribution of grain boundary diffusion[18], which leads to a modified equation:

$$L_{1D} = \sqrt{L_0^2 + 8\frac{p\Omega t}{k_B T}\left(\frac{\delta_I}{w}D_I\right)} \quad (2)$$

$$L_{2D} \approx \sqrt{L_0^2 + 4\frac{p\Omega t}{k_B T}\left(\frac{\delta_I}{w}D_I + \frac{\delta_{GB}}{d_{GB}}D_{GB}\right)} \approx \sqrt{L_0^2 + 4\frac{p\Omega t}{k_B T}D_{GB}\left(\frac{\delta_I}{w} + \frac{\delta_{GB}}{d_{GB}}\right)} \quad (3)$$

where $D_I$ is the interfacial diffusivity, $\delta_I$ is the interfacial diffusion thickness layer (~ 1 nm)[11], $w$ is either diameter of the wire or width of the trench, $\delta_{GB}$ is the grain boundary diffusion thickness layer, $d_{GB}$ is the grain size, and $D_{GB}$ is the grain boundary diffusivity. From 1D TMNM experiments, $D_I$ has been shown to be comparable to $D_{GB}$ for most metals[11], which leads to the simplification shown in Equation 3. Previous studies measuring $\delta_{GB}$ have varied widely with values ranging between 0.5 nm - 1 µm[19–21]. Using the successful molding conditions of nc-Cu and single crystal Cu, we estimate that $\delta_{GB}$ is ~20 nm. Thus, GB diffusion can dominate TMNM for small grain sizes and play a significant role in microcrystalline grain sizes as well. From Equation 3, we note there is a weak grain size dependence of $L \propto d_{GB}^{-\frac{1}{2}}$. Creep mechanisms[22], such as Navarro-Herring or Coble creep, have a stronger grain size relation, $L \propto d_{GB}^{-\alpha}$, where $\alpha$ is between 2-3. This does not match the experimental data since the pressure difference for molding was double between nc-Cu and µc-Cu while the grain size difference was approximately 10x.

Any appreciable grain boundary diffusion in polycrystal samples must occur simultaneously with grain boundary sliding to conserve volume, prevent void formation, and prevent grain overlap[22–26]. Importantly, grain boundary sliding conserves grain orientation since grains slide past one another rather than rotating or growing. There are two dominant grain boundary sliding mechanisms: Rachinger sliding[25] that leads to minimal change in grain size and shape and Lifshitz sliding[22] that leads to elongated, columnar grains due to the motion of vacancies perpendicular to the stress direction. Since we did not observe any change in grain size or any noticeable grain elongation (Figure 2), we predict that Rachinger sliding is the dominant mechanism.

In conclusion, we successfully show 2D TMNM over large length scales making it a viable technique for wafer-scale fabrication of 2D nanostructures at BEOL compatible conditions. Since TMNM is a material-agnostic technique, the same approach can be used to fabricate 2D nanostructures of a wide variety of compounds including other metals, alloys, intermetallics, and metallic glasses. The simultaneous control of both morphology and grain size during 2D TMNM can be useful for a wide array of applications where current nanofabrication techniques are the bottleneck for further advancements. In the context of interconnects, fabrication of high aspect ratio single crystal Cu nanostructures can greatly reduce resistivity by mitigating grain boundary scattering[27], which currently limits further interconnect scaling via Damascene process[28]. For catalysis, increasing grain boundary density in nanostructures has been shown to increase catalytic activity compared to pristine surfaces[29]. For mechanical metamaterials such as nanolattices, nanocrystalline grain boundaries act as boundaries for dislocation motion, leading to enhanced strength and toughness[30].

**Methods**

Si Trench Fabrication for 2D Mold

The following steps were taken to fabricate an 8-inch Si wafer with a 20 ~ 40 nm gap trenches with a 1 µm pitch: thermal oxidation for hardmask formation with a 200 nm gap (1 µm pitch), followed by a photolithography process with a KrF scanner (ASML, PAS 5500/700D), hardmask etching (Lam Research, EXELAN-HPT), photoresist stripping (PSK, DAS-2000), post cleaning, silicon trench etching (Lam Research, TCP-9400DFM), another round of post cleaning, and hard mask removal through hydrofluoric acid. Finally, low-pressure nitride deposition (Centrotherm, E1200) for gap shrinkage to the target gap of 20 ~ 40 nm. The etching conditions were as follows: a mixture of $CF_4$, Ar, and $O_2$ for the Bottom Anti-Reflective Coating (BARC) etching, $C_4F_8$, Ar, and $O_2$ for the oxide etching, and He, $SF_6$, and $O_2$ gases for silicon trench etching. After dicing into 7 mm x 7 mm sections, the diced wafers are cleaned in piranha solution before being transferred into the argon glovebox for TMNM.

Thermomechanical Nanomolding

Nanocrystalline 9 µm thick Cu foil (99.8% purity) was purchased from MTI. Microcrystalline 25 µm thick Cu foil (99.8% purity) was purchased from ThermoScientific. No additional annealing or polishing was done on the foils. Before TMNM, the Cu foil is cleaned in acetic acid for 30 seconds followed by blow-drying to remove the native oxide layer[31]. After transferring to an argon glove box, the Cu foil is placed on the Si wafer with trenches and the combination is placed into a steel die. All samples were first brought up to the final temperature at a ramp rate of 10°C/min before applying pressure using a hot press. After TMNM, the remaining foil is manually removed from the wafer using tweezers. No further processing or annealing is performed after molding.

For the single crystal sample, a pre-polished <110>-oriented Cu substrate was purchased from MTI. A Laue diffraction pattern was collected to determine the orientation of the (111) planes. From there the Cu substrate was rotated manually to orient (111) planes to be parallel to the trench walls before nanomolding. After TMNM, the Si was etched away using 15% KOH at 60°C followed by etching of the $Si_3N_4$ barrier layer using 5% HF.

SEM cross-section imaging and EBSD

SEM images were acquired using a Thermo Fisher Helios G4 UX Focused Ion Beam. EBSD maps were acquired using a Bruker eFlash HR EBSD in a Tescan Mira3 FESEM. The Cu foils were polished using a vibratory polisher with 70 nm colloidal silica for 3 hours for EBSD.

STEM imaging and 4D-STEM maps

Liftouts of Cu-filled trenches were made using a Thermo Fisher Helios G4 UX Focused Ion Beam with a final mill at 5 kV. STEM images were collected on a Thermo Fisher Scientific Spectra 300 S/TEM at 300 kV. The liftouts were initially tilted to align the Si on the <100>-axis, which is also the nanomolding direction. All 4D-STEM maps were collected on an Electron Microscope Pixel Array Detector (EMPAD). Grain mapping was done using Py4DSTEM[16]. A minimum of 7 Bragg diffraction spots were used to determine grain orientation.

**Acknowledgments**

MTK and JJC gratefully acknowledge support from NSF DMR under award 2240956. QPS is supported by an NSF GRFP under Grant Number 2139899. The authors acknowledge the use of facilities and instrumentation at Cornell Center for Materials Research (CCMR) supported by NSF Materials Research Science and Engineering Center DMR-1719875.